# Channel Boosted CNN-Transformer-based Multi-Level and Multi-Scale Nuclei Segmentation

*Zunaira Rauf[1, 2], Abdul Rehman Khan[1, 2], and Asifullah Khan[1, 2, 3*]*

[1]**Pattern Recognition Lab, Department of Computer & Information Sciences, Pakistan Institute of Engineering & Applied Sciences, Nilore, Islamabad 45650, Pakistan**

[2]**PIEAS Artificial Intelligence Center (PAIC), Pakistan Institute of Engineering & Applied Sciences, Nilore, Islamabad 45650, Pakistan**

[3]**Center for Mathematical Sciences, Pakistan Institute of Engineering & Applied Sciences, Nilore, Islamabad 45650, Pakistan**

**Corresponding Authors:** [*]**Asifullah Khan, asif@pieas.edu.pk**

## Abstract:

Accurate nuclei segmentation is an essential foundation for various applications in computational pathology, including cancer diagnosis and treatment planning. Even slight variations in nuclei representations can significantly impact these downstream tasks. However, achieving accurate segmentation remains challenging due to factors like clustered nuclei, high intra-class variability in size and shape, resemblance to other cells, and color or contrast variations between nuclei and background. Despite the extensive utilization of Convolutional Neural Networks (CNNs) in medical image segmentation, they may have trouble capturing long-range dependencies crucial for accurate nuclei delineation. Transformers address this limitation but might miss essential low-level features. To overcome these limitations, we utilized CNN-Transformer-based techniques for nuclei segmentation in H&E stained histology images. In this work, we proposed two CNN-Transformer architectures, Nuclei Hybrid Vision Transformer (NucleiHVT) and Channel Boosted Nuclei Hybrid Vision Transformer (CB-NucleiHVT), that leverage the strengths of both CNNs and Transformers to effectively learn nuclei boundaries in multi-organ histology images. The first architecture, NucleiHVT is inspired by the UNet architecture and incorporates the dual attention mechanism to capture both multi-level and multi-scale context effectively. The CB-NucleiHVT network, on the other hand, utilizes the concept of channel boosting to learn diverse feature spaces, enhancing the model's ability to distinguish subtle variations in nuclei characteristics. Detailed evaluation of two medical image segmentation datasets shows that the proposed architectures outperform existing CNN-based, Transformer-based, and hybrid methods. The proposed networks demonstrated effective results both in terms of quantitative metrics, and qualitative visual assessment.

# 1. Introduction

Nuclei segmentation plays crucial role in computational pathology, serving as a foundational step in the automated analysis of histological images. Among the various imaging modalities, the most commonly used staining is the Hematoxylin and Eosin (H&E) staining [1]. It allows for the quantification of features like nuclear size, shape, and intensity, which are crucial for tasks like cancer diagnosis, treatment planning, and disease prognosis [2]. However, accurate nuclei segmentation from H&E stained histology images presents considerable challenges due to several factors [3]. Firstly, nuclei often appear in clusters, making it difficult to discern individual nuclei boundaries. Secondly, there exists high intra-class variability among nuclei, both in terms of size and shape which further complicates the segmentation task. Additionally, nuclei may exhibit resemblance with other cellular structures, leading to poor segmentation results [4]. Moreover, the color and contrast correlation between nuclei and the background in H&E-stained images can hinder accurate segmentation [5].

In recent years, advancements in deep learning-based methods, especially convolutional neural networks (CNNs) have significantly contributed to medical imaging for tumor segmentation [6], [7], [8], cancer type classification [9] [10], nuclei segmentation [11], [12], and more [13], [14], [15]. Various architectures have been proposed to advance the research in the field of image segmentation [16], [17], [18]. Studies have shown that the incorporation of global semantic information in the segmentation models enhances their segmentation performance [19]. However, capturing long-range dependencies and global context information in the images for accurately delineating nuclei boundaries can be challenging for traditional CNNs [20]. To this end, many approaches have incorporated residual connections and attention mechanisms to retain spatial information [21], [22], but they still struggle to effectively integrate global-level contextual information. Recently, Transformer-based architectures have emerged as a promising alternative, leveraging self-attention mechanisms to capture global relationships within an image [23], [24], [25]. Nevertheless, Transformers may also face challenges in capturing fine-grained details and local features.

In this work, we carried out multi-level and multi-scale nuclei segmentation and proposed two hybrid architectures. Our proposed models utilize both CNNs and Transformer architectures to effectively delineate nuclei boundaries in histopathological images. In addition, we incorporated multi-axis attention and channel boosting to segment nuclei at multi-scale and multi-level. To deal with the complex nuclei representational variations we carried out some data processing to normalize the data (Fig.1 (a, b)). In this regard, we first proposed a UNet-like Nuclei Hybrid Vision Transformer "NucleiHVT" that employs convolutional and Transformer layers in each block, simultaneously learning enriched local and global features throughout the network (Fig. 1 (c)). The proposed NucleiHVT contains mobile convolution and multi-axis attention in its encoder and decoder architectures to capture local-global level features [26], [27]. Moreover, for enhanced nuclei segmentation, we proposed another architecture named Channel Boosted Nuclei Hybrid Vision Transformer "CB-NucleiHVT" (Fig.1 (e)). This architecture carries out channel boosting to learn enhanced feature representation by merging two feature extraction networks in the encoder block. We performed detailed experiments to analyze the effectiveness of the proposed models (Fig. 1 (d, f)). Both models showed reliable results on two multi-organ nuclei segmentation datasets as compared to other state-of-the-art techniques.

The key outcomes of our work are as follows:

1. Firstly, we present a stand-alone encoder-decoder hybrid vision Transformer architecture named NucleiHVT.

2. The proposed NucleiHVT-Encoder combines CNN layers and Transformer block in a deep sequential manner at each stage. It leverages a dual attention mechanism in the Transformer block to capture multi-level and multi-scale context effectively. In addition, the encoder alternates between conventional and dilated convolutions along the stages of the network to obtain features at different receptive fields.

3. The decoder architecture in the proposed NucleiHVT comprises deconvolution layers, followed by skip-connection, encoder-like processing blocks, and finally Transformer block with dual attention.

4. We also developed a Channel Boosted version of the Nuclei-HVT, "CB-NucleiHVT" that utilized the boosted channels from both the proposed NucleiHVT and the famous MaxViT architecture. In addition, we proposed a CB-Fusion-Block to effectively fuse diverse feature spaces from both architectures, thus enriching the output feature space.

5. Extensive experiments conducted on diverse medical image segmentation datasets showed the success of the proposed models.

The paper is organized as follows: Section 2 provides a brief overview of related works. Section 3 presents the detailed methodology of our developed techniques. In Section 4, we present the experimental results and discuss their implications. Finally, in Section 5, we summarize the main findings and conclude the paper.

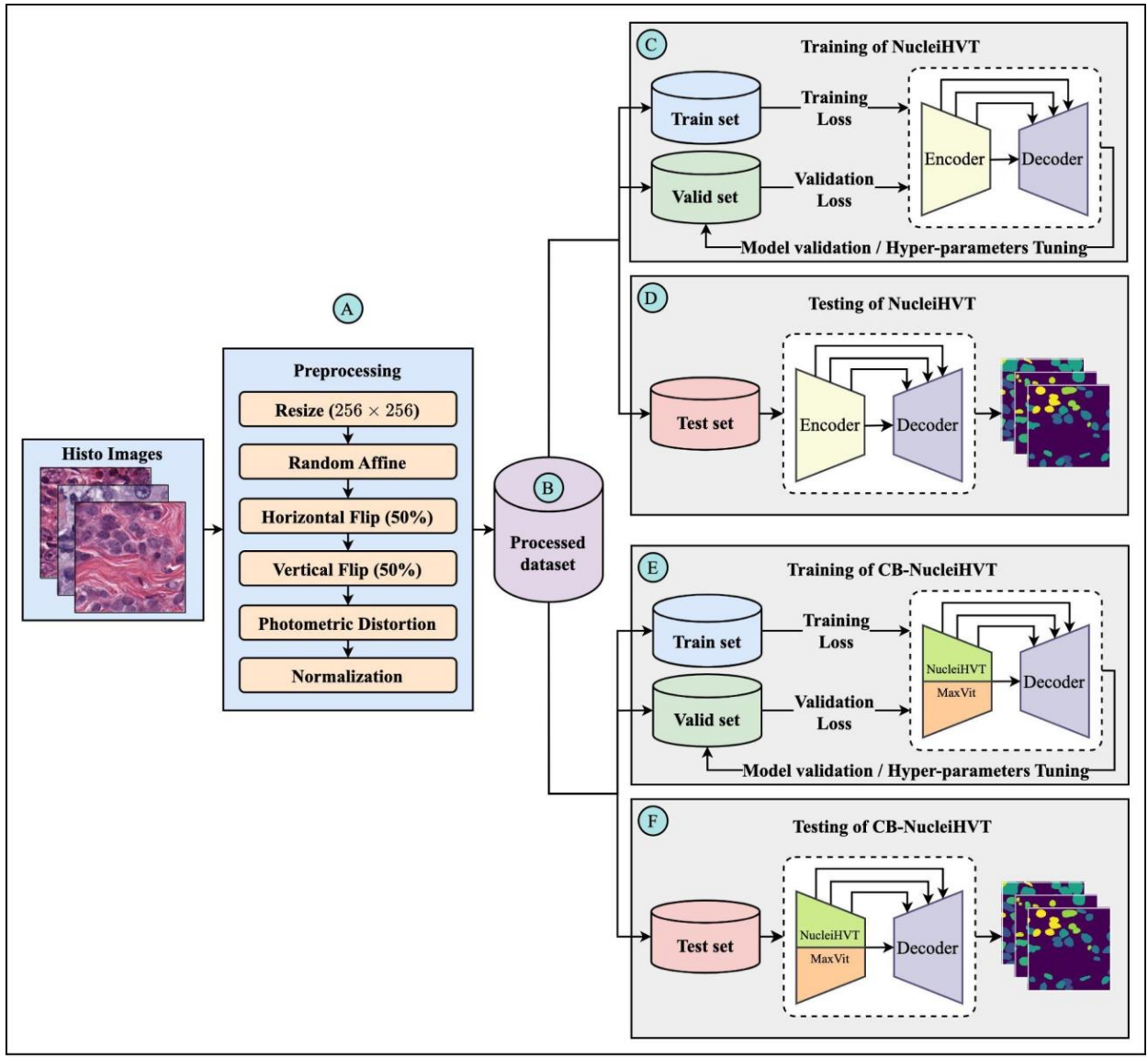

*Figure 1: Overall workflow of the Training and Testing pipelines for NucleiHVT and CB-NucleiHVT models.*

## 2. Related Work

Medical image segmentation requires a high-level understanding of the images due to the presence of multiple structures with variable morphology. Numerous approaches based on deep learning have been developed for segmentation in medical images. These methods tend to delineate the complex boundaries of various structures or regions by learning the hierarchical features and contextual information. Such approaches can be categorized into CNN-based, Transformer-based, or CNN-Transformer-based methods.

## 2.1 CNN-based Methods

CNN-based methods have been extensively utilized to solve complex computer vision problems ranging from medical images to natural images. These methods can effectively segment complicated structures in medical images due to their ability to learn and represent spatial hierarchies in the data. U-Net, proposed by Ronneberger et al., 2015 became famous due to its tremendous segmentation performance in medical images. Many approaches utilized U-Net-like models to carry out medical image segmentation [28], [29]. ResUNet [30] introduced the weighted attention and skip-connections in the encoder to achieve better results. Both the UNet++ [31] and UNet3+ [32], captured multi-scale features and balanced encoder/decoder semantic gap by utilizing dense connections in place of skip connections. GA-UNet [33] utilized a VGG-based UNet architecture to segment 2D and 3D medical images. Hoorali et al., 2022 proposed IRUNet [34] in which they utilized EfficientNet [35], ResNet [36], and Inception [37] layers into the UNet architecture to carry out multi-scale feature extraction. MultiResUNet [38] proposed MultiRes blocks that utilize multiple convolutional layers in a block for enhanced segmentation performance. Contour-aware Informative Aggregation Network (CIA-Net) [39] developed a new architecture that utilized a fusion strategy to fuse multi-level information between decoders. It aggregates task-specific features in a bi-directional fashion to capture spatial and texture correlations between nuclei and contour. Hover-Net [5] introduced the concept of horizontal/vertical distance maps to enhance boundary delineation. It also addresses the intra-class variability problem for different kinds of cells.

Despite the great progress of CNN-based medical image segmentation approaches, they may face difficulty in modeling global-level information, affecting their performance [6]. Even state-of-the-art models may face difficulty in capturing the target structures with significant inter-patient textural, and morphological variations [40]. To address this, numerous techniques have been utilized by the researchers to incorporate global information in the models. Various studies utilized skip connections in their encoder-decoder architectures for a smooth gradient flow. In addition, some approaches integrated attention mechanisms to focus on the relevant regions [41]. Axial-Deeplab factorized the 2D convolution into two 1D convolutions and proposed a module named position-sensitive gated axial-attention to rank the features [42]. Non-local U-Nets [43] proposed global aggregation blocks to model long-range semantics and overcome the limitations of local convolution operators. Spatial-Channel Attention U-Net (SCAU-Net) [44] integrates spatial and channel attention modules for local feature enhancement and restriction of irrelevant ones.

Another emerging approach for medical image segmentation is Channel Boosted CNNs (CB-CNNs) [45], [46], [47], which fuse diverse feature spaces from multiple encoders to enhance the quality of segmentation models. Channel-boosted techniques have shown exemplary performance in medical images across multiple modalities.

## 2.2 Transformer-based Methods

Recently, Transformers, have become a promising approach for image segmentation and other computer vision tasks. Due to their self-attention mechanism, they can model long-range relationships in the images, thus widely applied to medical images. Swin-Unet [48], proposed an Encoder-Decoder model that is inspired by the Swin Transformer [49], aiming to increase generalization and segmentation accuracy. Swin-Unet utilized shifted window-based Swin Transformer blocks as their basic unit. The nnFormer [50] incorporates specialized self-attention mechanisms within the Transformer blocks to capture the feature pyramids across the entire volume. MISSFormer [51], an encoder-decoder network with a hierarchical structure, also proposed a modified self-attention module to make it more computationally efficient. In addition, they also utilize an enhanced transformer context bridge to efficiently capture local and global context. TransDeepLab [52] combines the DeepLab and SwinTransformer. DeepLabv3 [53] effectively utilizes Atrous Spatial Pyramid Pooling (ASPP) and depth-wise separable convolution for multi-scale feature learning and reduced computational burden. By integrating Swin-Transformer modules with windows of different sizes, TransDeepLab efficiently fuses multiscale information within a lightweight model.

Even though Vision Transformers show exceptional performance in a variety of image segmentation tasks, they still face the problem of computational overload.

## 2.3 CNN-Transformer-based methods

Inspired by the potency of the convolutional operator to extract local features and the capability of the Transformer block to extract global features, many hybrid architectures have emerged recently. They broadly fall into the following sub-categories:

**Hybrid-Encoder Techniques:** TransUNet [40], a hybrid U-shaped architecture leverages the advantages of both U-Net and Transformer in the encoder, facilitating the recovery of localized spatial information and boosting segmentation performance. TransBTS [54], an encoder-decoder architecture combines a Transformer and a 3D CNN to learn both global and local image representations for medical image segmentation. Using the BiFusion module, TransFuse [55] explores the advantages of both CNNs and Transformers by combining CNN and Transformer modules. MedT [56] introduces an attention module that is position-sensitive, allowing the model to deal with multi-sized datasets. For 3D segmentation, UNETR [57] introduced a novel Transformer-based architecture that establishes a UNet-like connection between the Transformer-based encoded features and CNN-based decoded features at different resolutions. Swin UNETR [58], introduced a U-shaped architecture with a Transformer encoder and a CNN decoder for brain tumor segmentation. Swin UNETR uses a shifted windowing mechanism for self-attention computation.

**Hybrid-Decoder Techniques:** The Segtran framework [59] proposes the Squeeze-and-Expansion Transformer for image segmentation, with a Learnable Sinusoidal Position Encoding to introduce continuity bias. The architecture includes a CNN backbone for feature extraction, input/output feature pyramids for upsampling, and Squeeze-and-Expansion Transformer layers for contextualization. A novel Transformer decoder architecture combines Squeezed Attention Blocks (SAB) with Expanded Attention Blocks (EAB).

**Hybrid-Encoder-Decoder Techniques:** Recently, MaxViT-UNet [60] introduced a novel CNN-Transformer approach to segment medical images. It utilized a hybrid encoder that contained both CNN and Transformer blocks [72]. The proposed decoder, also inspired by the MaxViT's multi-axis attention [27] (Max-SA), utilizes Max-SA to capture local and global features at each level, essential for accurate medical image segmentation. By focusing on features along several axes, the Multi-Axis Attention mechanism improves the ability to distinguish between background and object regions, resulting in efficient segmentation.

The hybrid approaches may still face issues due to the self-attention mechanism's quadratic nature and the effective fusion of CNNs' and Transformers' local and global feature extraction capabilities. Thus, in this work, we developed that effectively exploited both convolution and self-attention at each step. Furthermore, our method is fast and reliable for medical image segmentation tasks since it uses a linear multi-axis attention mechanism.

## 3. Materials and Methods

### 3.1 Dataset Description

**MoNuSeg18**

The MoNuSeg (2018) provided a binary nuclei segmentation challenge dataset for histopathology images [61]. The images were obtained from 18 different medical centers with their own specific staining techniques and image acquisition techniques. They contained images from seven different organs, i.e. breast, colon, liver, bladder, stomach, kidney, and prostate. This diversity of sources and organs ensured that the dataset contained various nuclear shapes and textures. The MoNuSeg 2018 challenge dataset consisted of 30 training images and 14 test images, each having a resolution of 1000×1000 and scanned at a 40× magnification. The train set comprised 21,623 manually annotated nuclei boundaries, whereas the test set contained around 7,233 annotated nuclei boundaries. The test set was more challenging by including two new organ types, the lung and brain, and excluding the stomach and prostate in the train set.

**MoNuSAC20**

The MoNuSAC20 (2020) provided a multi-class nuclei segmentation challenge dataset for histopathology images [62]. It was designed to represent different organ types and nuclear categories and comprised the Epithelial, Lymphocytes, Macrophages, and Neutrophils nuclear types. The train set was prepared by cropping the Whole slide images (WSIs), taken from the TCGA data portal [63]. These WSI were obtained from 32 various medical centers and 46 distinct patients and were captured at 40× magnification. Every nucleus in the dataset has annotations at the class and boundary levels.

The dataset provides boundary-level and class-level annotations for each nuclei. The test set was prepared by collecting the samples from 25 patient samples, and these samples were collected from 19 various hospitals, of which 14 hospitals overlapped with the train set. Similar data preparation steps were performed for the testing data, however annotations for unclear areas were also included. These areas include regions with vague nuclei, unclear boundaries, or ambiguous labels.

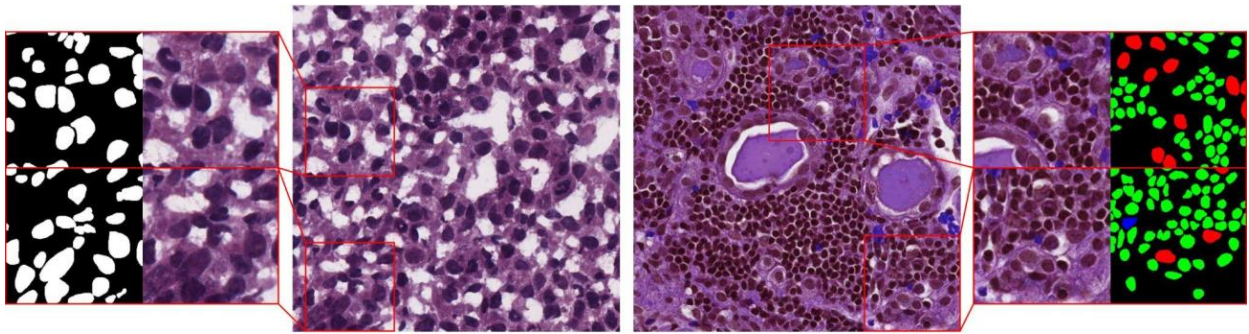

*Figure 2: Hematoxylin and Eosin stained whole slide images and their respective patches and ground truths. Figures A and B are the whole-slide images from MoNuSeg and MoNuSAC respectively.*

### 3.1.2 Dataset Preparation

**MoNuSeg18**

For the training and testing of the segmentation models, we prepared the MoNuSeg18 dataset [61] by performing some preprocessing. Firstly, we obtained 256x256-sized patches from the provided 1000x1000 images. In addition, we applied several augmentation techniques to make the dataset more robust, such as RandomAffine, PhotoMetricDistortion, Random Horizontal, and Vertical Flip with a 0.5 flip probability. We also applied mean and standard deviation-based normalization on the training and testing sets.

**MoNuSAC20**

For MoNuSAC20 dataset [62], we carried out the same pre-processing techniques as that of MoNuSeg18. We also applied data augmentation, and data normalization in addition to data resizing.

## 3.2 Proposed NucleiHVT

The proposed NucleiHVT is a U-Net-like architecture with a proposed encoder and decoder. Figure 2 depicts the network architecture of the developed NucleiHVT. The developed network has symmetric encoding and decoding blocks throughout its architecture. These blocks leverage both CNN and Transformer layers to learn local as well as global dependencies in the images. In addition, we also utilized residual connections among the encoder and decoder architectures for a smooth gradient flow and to capture contextually rich features, thereby enhancing the proposed NucleiHVT's segmentation ability.

### 3.2.1 NucleiHVT-Encoder

Inspired by the recent success of hybrid vision Transformers [64], [65] in several image segmentation tasks, the developed NucleiHVT employs a hybrid encoder architecture. The proposed NucleiHVT encoder initially contains two CNN transformation layers in the first stage ($S_0$) followed by three encoding stages ($S_1$-$S_3$) and one bottleneck stage ($S_4$). Each encoding stage (S0-S4) contains several CNN layers and a MaxViT block. The CNN layers use multiple 1x1 and 3x3 filters to capture the domain-relevant features. The MaxViT block [27] utilizes Mobile Convolution block MBConv [26] and multi-axis attention [27] mechanism not only for computationally efficient feature extraction but also to model the global-level information. In the encoding stage S2 and S4, we also utilized dilated convolutions along with conventional convolutions to capture the multi-scale context while preserving the spatial resolution. Dilated convolutions provide larger receptive fields without increasing parameters which makes them computationally efficient. CNN layers in all the stages are followed by Batch Normalization and Mish non-linearity layers. We also incorporated residual connections to alleviate the problems like vanishing or exploding gradients. Moreover, the Maxpool layer was also utilized to preserve the most important features and to reduce the spatial size of the feature maps. Our proposed encoder incorporates multiple architectural modifications in each stage for robust and effective multi-scale feature extraction.

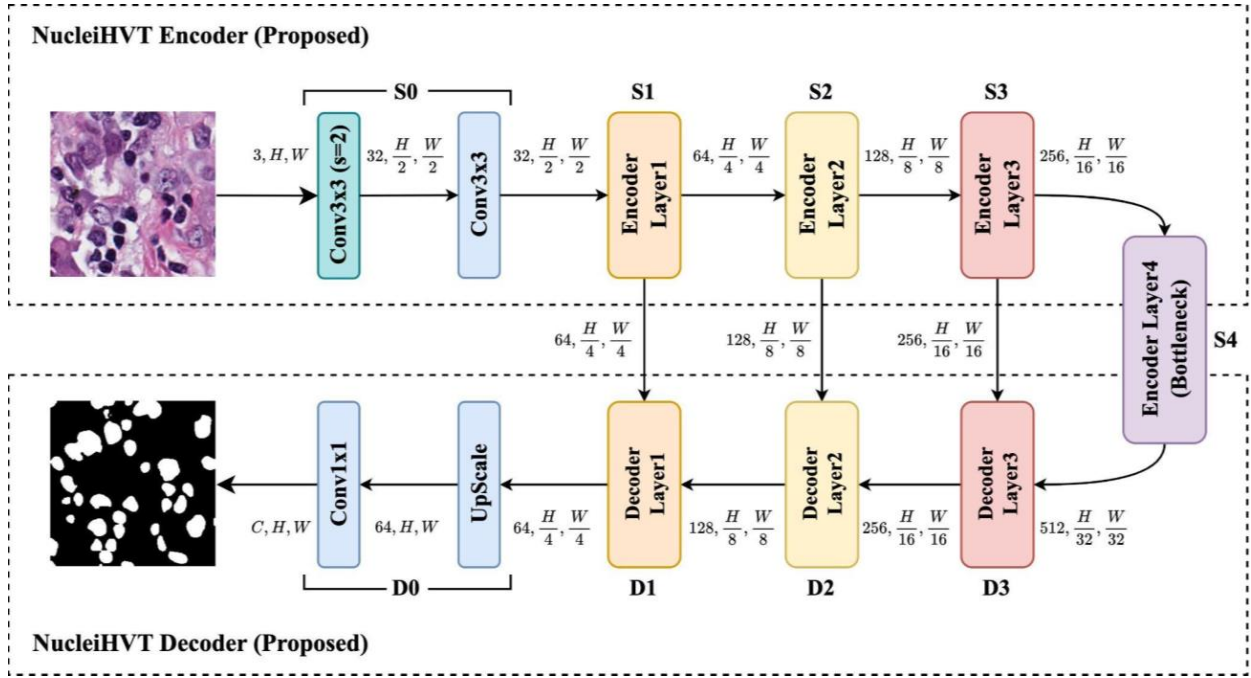

*Figure 3: Overview of the Proposed NucleiHVT-Net framework. The proposed architecture follows a UNet-like structure, containing several encoding and decoding stages.*

### 3.2.2 NucleiHVT-Decoder

The proposed NucleiHVT-Decoder also has a similar architecture as the Nuclei-HVT encoder with several decoding stages (D3-D0). Like the NucleiHVT-encoder, the decoder is also a hybrid vision Transformer architecture with both CNN and multi-attention layers. After the bottleneck block (S4) the contextual rich feature maps are fed to the first decoding stage D3, where the deconvolution layer is employed to increase the spatial dimensions of the feature maps. The skip connections were utilized in the architecture from each encoding stage to its decoding stage for an efficient gradient flow, to retain the high-resolution feature and to prevent overfitting. Semantically and spatially rich features were obtained in each decoding stage by concatenating the up-sampled features with the relevant skip-connection features after the deconvolution layer. A series of convolution layers are then utilized followed by max pooling and MaxViT block for an effective and accurate reconstruction of output semantic masks.

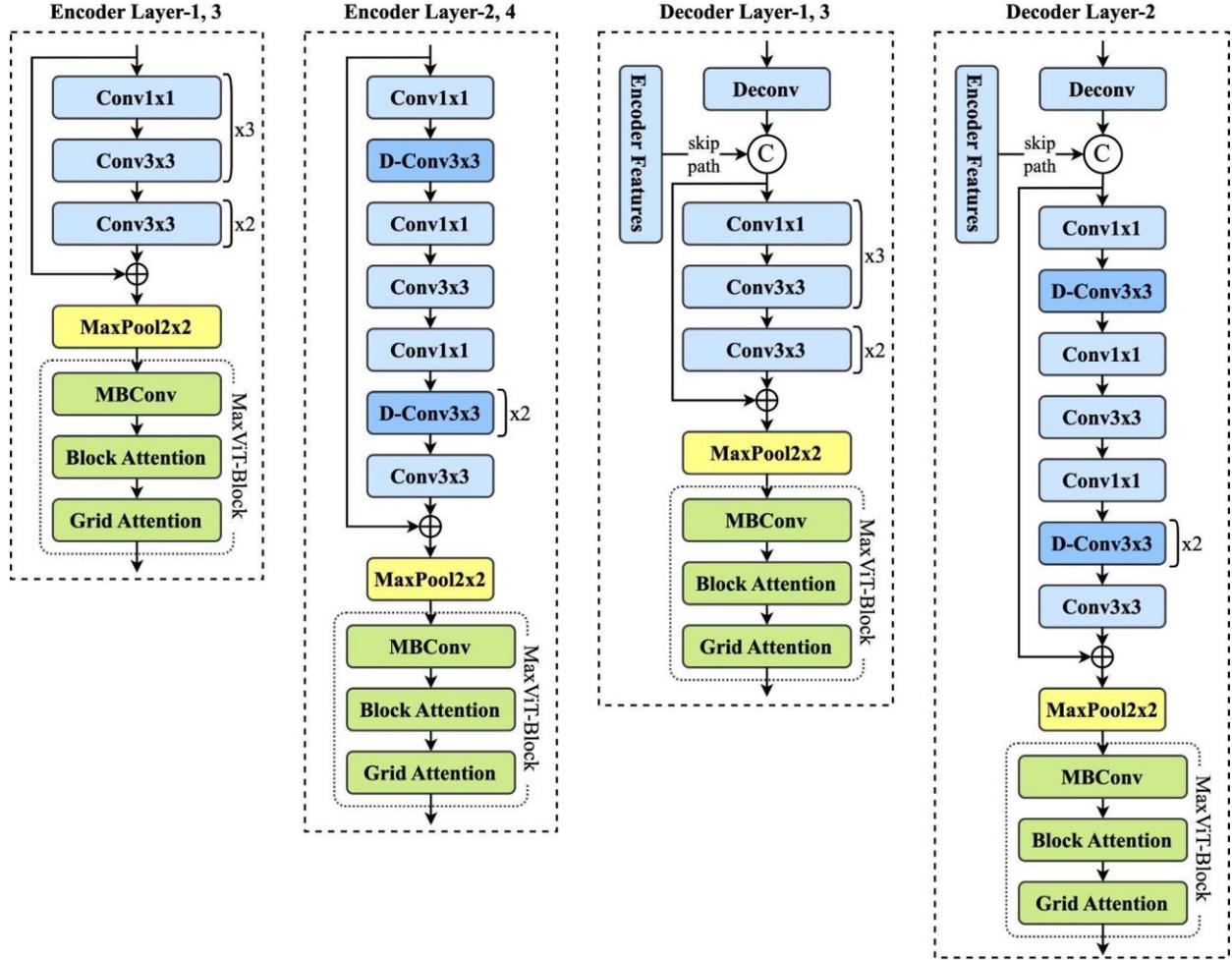

*Figure 4: Detailed architecture of the proposed Encoder and Decoder blocks. The left column shows encoder blocks, whereas the right column shows decoder blocks. The hybrid nature of each block allows effective multi-scale feature extraction.*

## 3.3 Proposed CB-NucleiHVT

Medical images exhibit high-level pattern diversity, therefore at times, a single learning model may result in limited performance. In this regard, we employed the idea of channel boosting to develop CB-NucleiHVT to segment nuclei in medical images. In channel boosting, the feature spaces from multiple models are extracted and then fused efficiently to obtain a merged enriched feature space. The channel boosting mechanism relies on an efficient fusion mechanism to retain relevant features from multiple spaces and suppress redundant and noisy features. In the proposed CB-NucleiHVT, the encoder architecture utilized channel-boosted feature space obtained from MaxViT-UNet and the proposed NucleiHVT. The hierarchical features from both backbones are fused level-wise using the novel CB-Fusion block. The CB-Fusion block leverages block and grid attention mechanisms for local and global processing respectively, providing spatially enriched feature maps. The proposed CB-NucleiHVT utilizes the same decoder as the proposed NucleiHVT to decode the encodings learnt from the encoder architecture. Both the encoders and the decoder architectures were initialized using weights pre-trained on the MoNuSeg18 and MoNuSAC20 datasets individually to obtain better weight initialization.

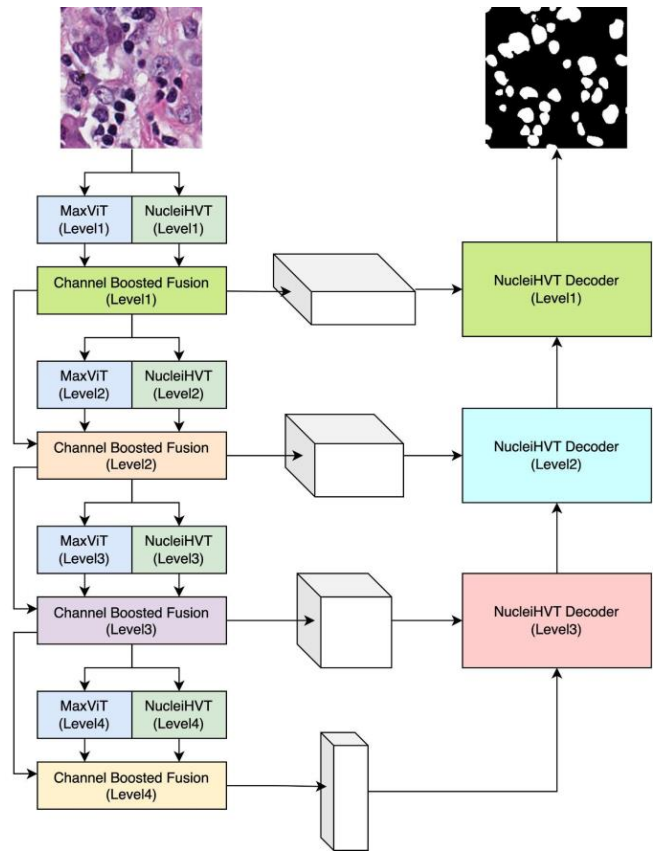

*Figure 5: Architectural diagram of the proposed CB-NucleiHVT.*

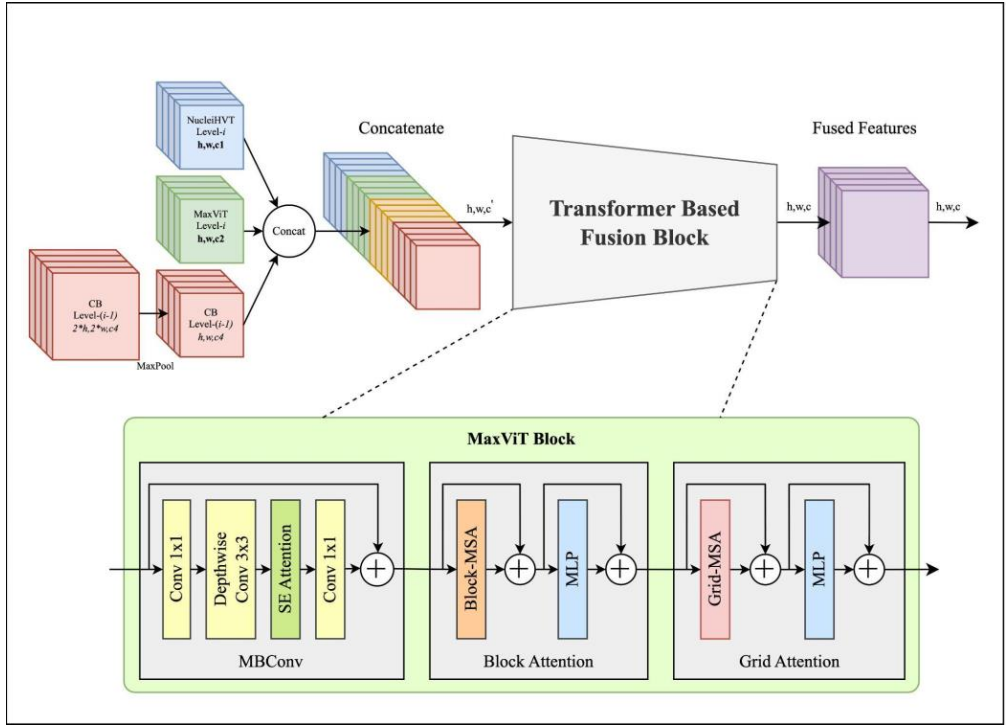

*Figure 6: Architecture of the proposed CB-Fusion block. The features from multiple feature extractors are merged using a hybrid fusion block, which allows for local and global feature processing for better feature refinement.*

## 3.4 Loss function of the proposed approaches

The loss function used for training both the proposed NucleiHVT and CB-NucleiHVT consisted of a weighted combination of CrossEntropy and Dice loss functions. Dice and cross-entropy losses are computed pixel-by-pixel, but the former loss focuses on individual pixels only, while the later loss also takes account of neighborhood pixels due to the intersection and union calculations involved. The combined loss function is mathematically formulated as follows:

$$Loss(y_i, \hat{y}_i) = w1 * CrossEntropyLoss(y_i, \hat{y}_i) + w2 * DiceLoss(y_i, \hat{y}_i) \quad (5)$$

$$CrossEntropyLoss(y_i, \hat{y}_i) = -\sum_{i=1}^{H*W} y_i * log(\hat{y}_i) \quad (6)$$

$$DiceLoss(y_i, \hat{y}_i) = -2 * \frac{|y_i \cup \hat{y}_i|}{|y_i \cap \hat{y}_i|} \quad (7)$$

Where we used the weight values of 1 and 3 for w1 and w2 respectively for both challenge datasets.

## 3.5 Training and implementation details

All the experiments were done on the NVIDIA DGX Station with 4 Tesla V100 GPUs, 120GB GPU memory, 256 GB RAM, and an Intel Xeon E5-2698 CPU. The PyTorch (v1.12.1) based framework named MMSegmentation [66] (v0.24.1) was utilized for experimentation using Conda

(v4.12.0) environment setup. For rapid experimentation, we utilized the distributed training provided by MMSegmentation. The batch size for a single GPU was set to 4, with an effective batch size of 16 images on 4 GPUs. We experimented with SGD, Adam [67], AdaBelief [68], and AdamW [69] optimizers, and found the results of AdamW optimal among them. For the results reported in the next section, AdamW was used with a learning rate of 0.005 for NucleiHVT and 0.001 for CB-NucleiHVT-UNet, the weight decay was set to 0.001, and betas were set to (0.9, 0.999) in AdamW. We utilized CosineLRScheduler for gradually decreasing the learning rate during model training.

### 3.6 Evaluation Metrics

The commonly used evaluation metrics for segmentation-based approaches are Dice and IoU. In case of image segmentation, Dice is the same as the F1-Score, and IoU is also known as the Jaccard Index. To evaluate both NucleiHVT and CB-NucleiHVT on segmentation challenges, MoNuSeg18 and MoNuSAC20, the Dice and IoU evaluation metrics were calculated for all experiments. IoU and Dice are mathematically defined in the equations 8 and 9 shown below.

$$IoU(Y_t, \hat{Y}_p) = \frac{|Y_t \cap \hat{Y}_p|}{|Y_t \cup \hat{Y}_p|} \qquad (8)$$

$$Dice(Y_t, \hat{Y}_p) = \frac{2.|Y_t \cap \hat{Y}_p|}{|Y_t|+|\hat{Y}_p|} \qquad (9)$$

where $Y_t$ is the true mask image and $\hat{Y}_p$ is the predicted mask image.

### 3.7 Comparative Study

The proposed NucleiHVT and CB-NucleiHVT are compared with previously existing techniques on both the challenge datasets. The comparison is specially made with CNN-based and ViT-based models to highlight the effectiveness of the hybrid nature of the proposed techniques. The CNN-based technique is a vanilla UNet model, whereas the ViT-based model chosen for comparison is the state-of-the-art Swin-UNet framework. Some of the recent techniques on MoNuSeg18 and MoNuSAC20 datasets are also mentioned in the next section for better comparison across a range of models and techniques.

## 4. Experimental Results and Discussion

### 4.1. Quantitative results of the proposed approach

The comparative outcomes of the developed models, NucleiHVT and CB-NucleiHVT, are detailed in Tables 1 and 2 for the MoNuSeg18 and MoNuSAC20 datasets, respectively. These results are compared with those of previous methodologies. For the assessment of the proposed models on both datasets, we utilized MMSegmentation framework for training the proposed and existing models with the identical set of hyper-parameters. For we performed binary-class segmentation for the MoNuSeg18 challenge dataset, whereas for the MoNuSAC20 dataset which encompasses four distinct nuclei types, multi-class segmentation was done.

The NucleiHVT framework proposed in this study outperforms previous techniques significantly on both datasets, highlighting the significance of the CNN-Transformer architecture. Specifically, for the MoNuSeg18 dataset, the proposed NucleiHVT surpasses the CNN-only UNet [70] with the 3.41% IoU score and 2.70% Dice score, as well as the Transformer-only Swin-UNet [48] by 10.69% in IoU score and 5.66% in Dice score. Whereas, in the case of MoNuSAC20 dataset, the developed NucleiHVT outperforms UNet by 14.67% in IoU score and 10.99% in Dice scores, surpassing Swin-UNet by significant values in both Dice and IoU metrics, as illustrated in Table 2.

The channel-boosted proposed framework CB-NucleiHVT outperforms not only the previous techniques significantly on both datasets but also gives better results than NucleiHVT underscoring the importance of the diverse feature extractors and the concept of channel boosting. Specifically, it surpasses the CNN-only UNet by 5.11% in IoU score and 2.94% in Dice score, as well as the Transformer-only Swin-UNet by 12.52% in IoU score and 5.91% in Dice score on the MoNuSeg18 dataset. Compared with the first proposed technique, NucleiHVT, it improved the IoU and Dice by 1.65% and 0.24%, respectively. In the case of the MoNuSAC20 dataset, the proposed NucleiHVT outperforms UNet with 17.60% in the IoU score and 12.91% in the Dice scores. NucleiHVT also showed reliable results in terms of both Dice and IoU metrics when compared with Swin-UNet, as illustrated in Table 2. It is also evident that the channel-boosted technique resulted in 2.55% and 1.73% improvements in IoU and Dice metrics over NucleiHVT. The substantial improvements in both Dice and IoU scores depict the significance of the hybrid encoder-decoder architectures and the idea of channel boosting.

*Table 1: Comparative results on MoNuSeg18 Dataset (nucleus vs background)*

| Method | IoU | Dice |
|---|---|---|
| U-Net [70] | 0.6927 | 0.8185 |
| U-net++ [31] | 0.6089 | 0.7528 |
| Bio-net [71] | 0.6252 | 0.7655 |
| ATTransUNet [72] | 0.6551 | 0.7916 |
| TransUnet [40] | 0.6568 | 0.7920 |
| MultiResUnet [38] | 0.6380 | 0.7754 |
| MBUTransNet [73] | 0.6902 | 0.8160 |
| UCTransnet [74] | 0.6668 | 0.7987 |
| FSA-Net [75] | 0.6699 | 0.8032 |
| MedT [56] | 0.6573 | 0.7924 |
| AttentionUnet [76] | 0.6264 | 0.7620 |
| Swin-Unet [48] | 0.6471 | 0.7956 |
| NucleiHVT (Proposed) | **0.7245** | **0.8402** |
| CB-NucleiHVT (Proposed) | **0.7281** | **0.8426** |

*Table 2: Comparative results on MoNuSAC20 Dataset (nucleus vs background)*

| **Method** | **IoU** | **Dice** |
|---|---|---|
| UNet [70] | 0.5874 | 0.7197 |
| MulVerNet [77] | - | 0.7660 |
| NAS-SCAM [78] | - | 0.6501 |
| Hover-net [5] | - | 0.7626 |
| Dilated Hover-net w/o ASPP [5] | - | 0.7571 |
| Dilated Hover-net w/ ASPP [5] | - | 0.7718 |
| Swin-Unet [4] | 0.3924 | 0.4689 |
| NucleiHVT-UNet (Proposed) | **0.6758** | **0.8008** |
| CB-NucleiHVT-UNet (Proposed) | **0.6908** | **0.8126** |

*Table 3: Comparison of model parameters and flops between proposed models and baselines.*

| **Model** | **Parameters (Million)** | **Flops (GFLOPs)** |
|---|---|---|
| Unet | 29.06 | 50.64 |
| Swin-UNet | 27.29 | 11.31 |
| NucleiHVT | 33.46 | 18.99 |
| CB-NucleiHVT | 65.37 | 25.57 |

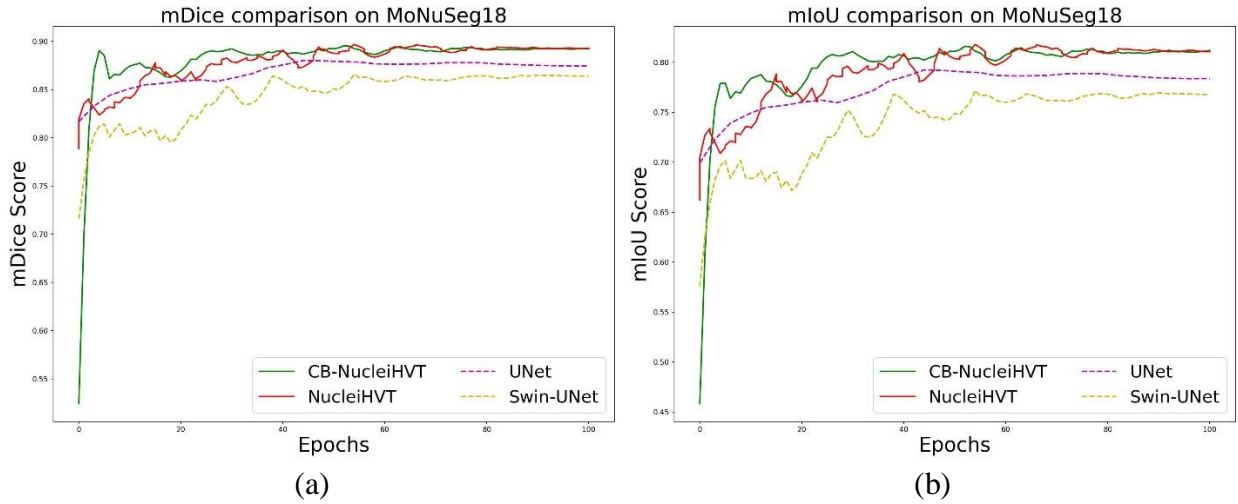

*Figure 7: Comparison of (a) mDice (mean Dice) and (b) mIoU (mean IoU) metrics between proposed models (CB-NucleiHVT (green) and NucleiHVT (red)) and baseline models (UNet (purple) and Swin-UNet (yellow)) on the MoNuSeg18 dataset.*

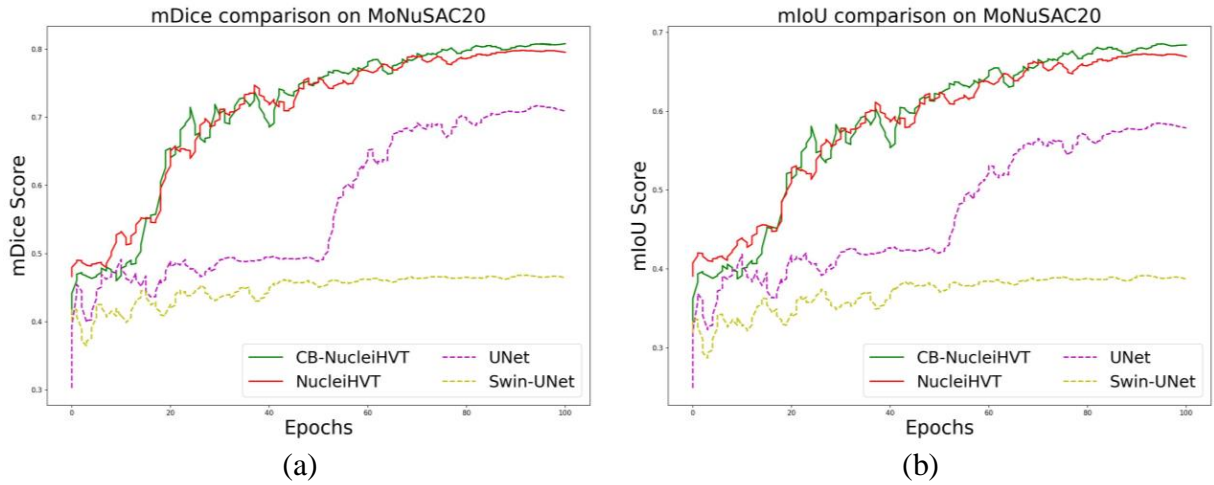

*Figure 8: Comparison of (a) mDice (mean Dice) and (b) mIoU (mean IoU) metrics between proposed models (CB-NucleiHVT (green) and NucleiHVT (red)) and baseline models (UNet (purple) and Swin-UNet (yellow)) on the MoNuSAC20 dataset.*

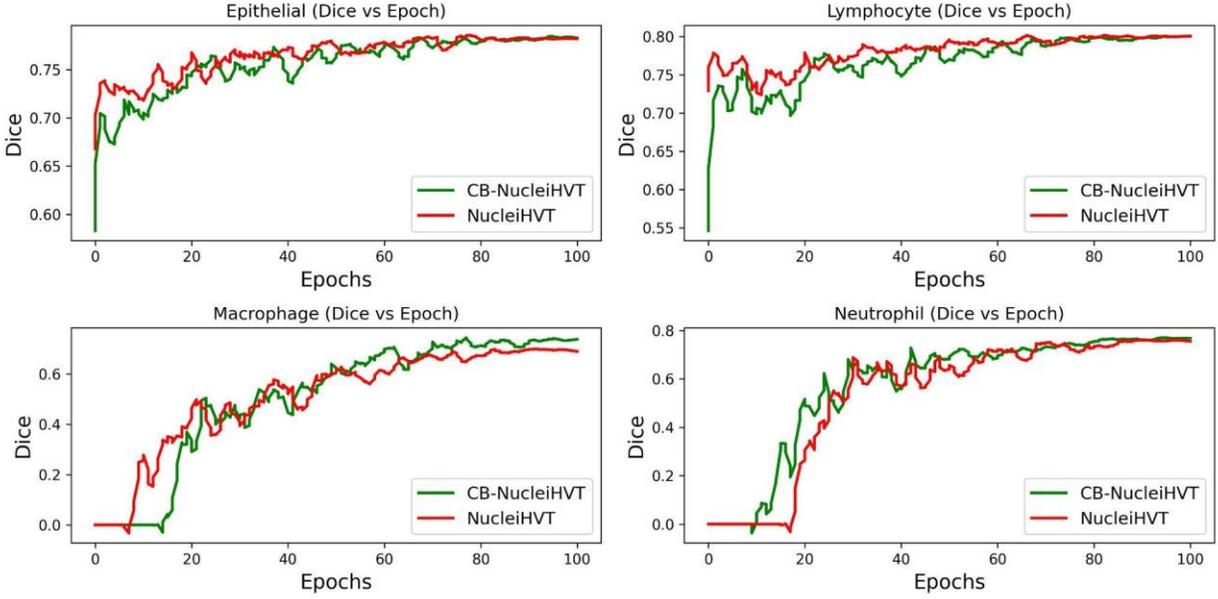

*Figure 9: Class-wise comparison of Dice metric between proposed CB-NucleiHVT (green) and NucleiHVT (red) on the MoNuSAC20 dataset. The CB-NucleiHVT performed remarkably well on all four classes due to the fusion of hybrid feature spaces.*

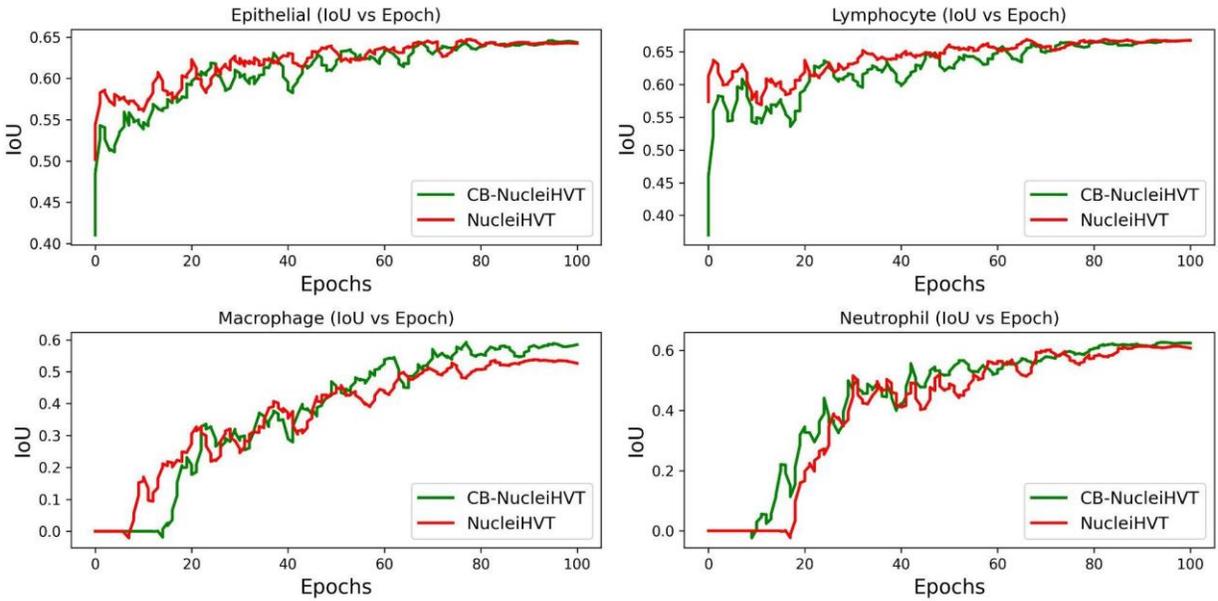

*Figure 10: Class-wise comparison of IoU metric between proposed CB-NucleiHVT (green) and NucleiHVT (red) on the MoNuSAC20 dataset. The CB-NucleiHVT performed remarkably well on all four classes due to the fusion of hybrid feature spaces.*

### 4.2. Qualitative results of the proposed approach

The qualitative assessment on various image samples from both the MoNuSeg18 (Figure 4a) and MoNuSAC20 datasets (Figure 4b) exhibit the efficacy of NucleiHVT and CB-NucleiHVT techniques. The generated masks of proposed techniques exhibit lower error susceptibility in comparison to Swin-UNet [48] and vanilla UNet [79], while also providing masks with comparatively precise boundaries. Further visual comparisons between ground truth and predicted mask images are presented in Figures 5a and 5b, where white indicates accurately predicted regions, and colored regions highlight erroneous predictions. In Figure 5b, we depicted diverse nuclei classes with variable colors in the MoNuSAC20 dataset.

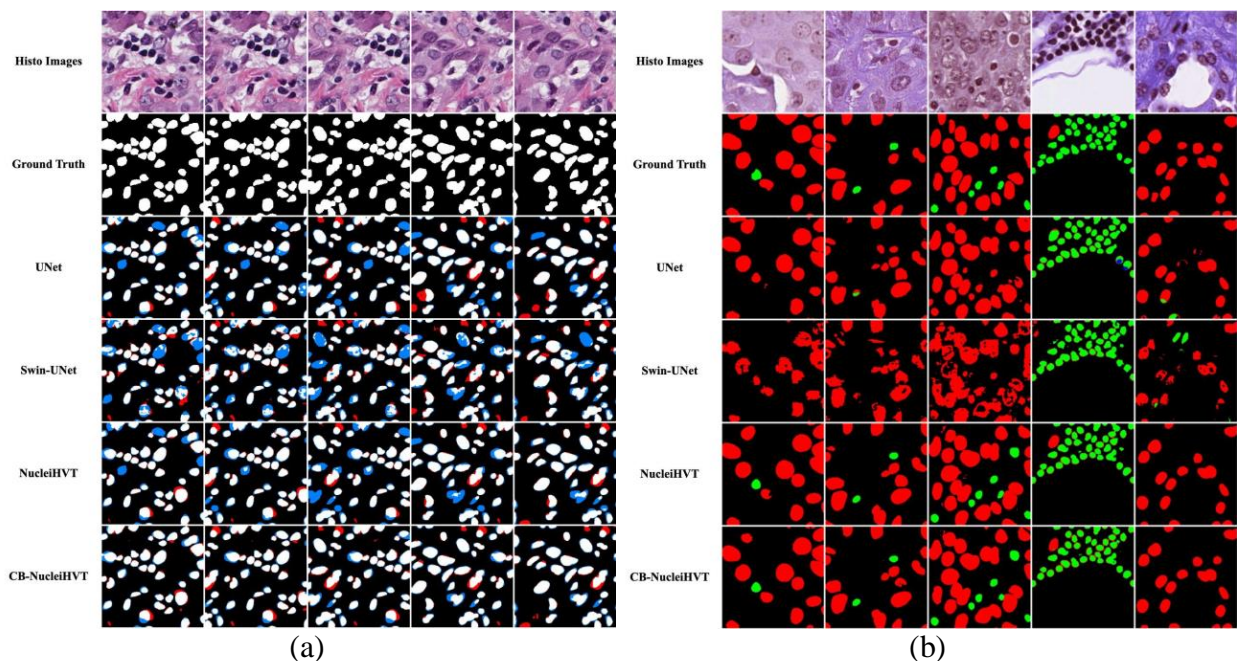

Figure 11: Output comparison of the proposed NucleiHVT and CB-NucleiHVT on MoNuSeg18 dataset (left) and MoNuSAC20 dataset (right).

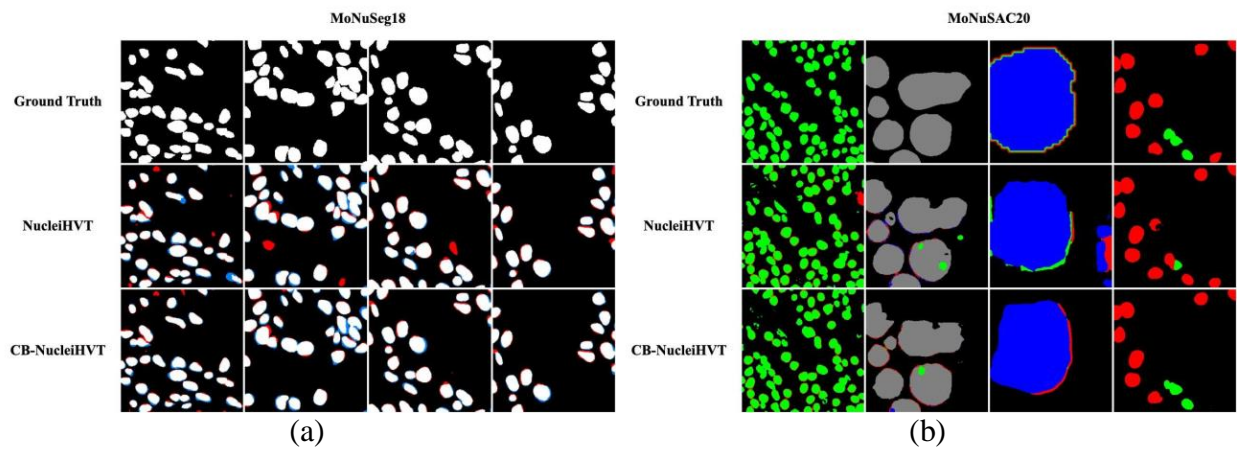

Figure 12: Output comparison of the proposed NucleiHVT and CB-NucleiHVT on MoNuSeg18 dataset (left) and MoNuSAC20 dataset (right).

## 4.3. Ablation Study

In this section, we present the comparison of the proposed hybrid NucleiHVT Decoder with other CNN-based and hybrid decoders. For the CNN decoder, we chose the UPerNet decoder which was proposed in [80]. The UPerNet decoder is purely CNN in nature and relies on Pyramid Spatial for dense feature extraction. For comparison with the hybrid decoder, we choose the recently proposed MaxViT-UNet decoder, which also incorporates the hybrid attention block in each decoding stage. The proposed decoder beat these previous decoding techniques by an optimal margin proving the superiority of our proposed decoding strategy. The results of our comparison are shown in Table 3.

*Table 4: Ablation results comparison of NucleiHVT Decoder with other decoders on MoNuSeg18 and MoNuSAC20 datasets*

| Encoder | Decoder | IoU | Dice |
|---|---|---|---|
| MoNuSeg18 | | | |
| NucleiHVT | NucleiHVT | **0.7245** | **0.8402** |
| NucleiHVT | MaxViT-UNet [60] | 0.7133 | 0.8327 |
| NucleiHVT | UPerNet [80] | 0.7163 | 0.8347 |
| MoNuSAC20 | | | |
| NucleiHVT | NucleiHVT | **0.6758** | **0.8008** |
| NucleiHVT | MaxViT-UNet [60] | 0.6736 | 0.7988 |

## 5. Conclusion

In this paper, we present two encoder-decoder CNN-Transformer architectures for medical image segmentation in multi-organ histology images. Firstly, we present a Nuclei Hybrid Vision Transformer (NucleiHVT) that leverages dual attention mechanisms and convolutional transformations to capture multi-level and multi-scale contexts effectively. The dual attention mechanism helped in capturing intricate patterns both locally and globally by utilizing its blocked local and dilated global attention mechanisms. Secondly, taking advantage of diverse feature spaces, we proposed a channel-boosted approach named CB-NucleiHVT. The encoder of the proposed framework exploits diverse feature learners to learn multi-variate nuclei features. Moreover, we employed a Channel Boosted Fusion Block to effectively merge the discriminating features and ignore the redundant ones. Extensive experiments conducted on diverse medical image segmentation datasets demonstrated the outperformance of the both NucleiHVT and CB-NucleiHVT techniques compared to other existing approaches. In future, we intend to find the effectiveness of the proposed architectures on other 2D/3D imaging modalities and real-life datasets. In addition, we will also exploit Self-Supervised Learning based pre-trained models in the Channel Boosted framework to develop more robust and generalized segmentation techniques.

## Acknowledgments

This work has been conducted at Pattern Recognition Lab (PRLab), Pakistan Institute of Engineering and Applied Sciences, Islamabad, Pakistan. We acknowledge the Pakistan Institute of Engineering and Applied Sciences (PIEAS) for a healthy research environment which led to the research work presented in this article.

## Authors Contributions

### Competing interests

The authors declare no competing financial and/or non-financial interests about the described work.

### Additional information

Correspondence and requests for materials should be addressed to A.K.